\documentstyle[12pt]{article}
\begin{document}
\centerline{\huge{\bf Critical exponents for}}
\centerline{\huge{\bf nuclear multifragmentation:}} 
\centerline{\huge{\bf dynamical lattice model}}
\bigskip

\centerline{\large{\bf J.S. S\'a Martins \footnote{Present address: {\it CIRES and Colorado Center for Chaos and Complexity, University of Colorado, Boulder, CO, 80309}} and P.M.C. de Oliveira}}

\centerline{\it Instituto de F\'{\i}sica, Universidade Federal Fluminense}

\centerline{\it Av. Litor\^anea s/n, Boa Viagem, Niter\'oi 24210-340, RJ, 
Brazil}

\centerline{\it e-mail: jssm@if.uff.br}

\bigskip
\noindent{\bf Abstract}
\bigskip

We present a dynamical and dissipative lattice model, designed to mimic nuclear multifragmentation. Monte Carlo simulations with this model show clear signature of critical behaviour and reproduce experimentally observed correlations. In particular, using techniques devised for finite systems, we could obtain two of its critical exponents, whose values are in agreement with those of the universality class to which nuclear multifragmentation is supposed to belong.\\



\bigskip
\noindent{\bf 1. Introduction}
\bigskip

Nuclear fragmentation is one of the most active lines of research in
contemporary nuclear physics and has been extensively investigated in
nucleus-nucleus and proton-nucleus collisions at intermediate and high
energies. Recent progress in experimental studies has produced
considerable evidence concerning its statistical properties.\\

The observation of a power law for the quantity of fragments, as a function of fragment size (mass or charge), produced in certain ranges of beam energies in these collisions, supported by the characteristic van der Waals behaviour of the nucleon-nucleon force, has stimulated the emergence of the idea of a liquid-gas phase transition in nuclear matter \cite{mekjian}. Its phase diagram has a line of first-order transitions ending at a critical point. At this critical point nuclear matter suffers a continuous transition, and the analog for the "critical opalescence" phenomenon \cite{stanley} is the fragment distribution's characteristic power law shape \cite{fisher}. From this perspective, nuclear multifragmentation can be considered as the remnant, for a finite system, of the proper transition which would happen in the thermodynamic limit.\\

The experimental search for signs of a nuclear matter phase transition faces at least two inherent - albeit not specific - complications:  

\noindent (1) Nuclei are finite systems, composed of a limited
number of constituents; for these systems, which do not suffer 
phase transitions in a strict sense, the divergences which characterize
the critical point are considerably broadened.  

\noindent (2) Nuclei are closed systems, and are not embedded in an
infinite heat bath; this fact prevents the predetermination of their
temperature, which has to be reconstructed from observable quantities -
the whole concept of temperature is, in this case, somewhat delicate.\\

Considerable effort has been made to develop a number of field-theoretical
models for infinite nuclear matter that yield such a transition, and some
success has been obtained in the application of these ideas to real
systems of finite nuclei \cite{matter,finite}.\\

However, within the context of those models, the analytical derivation of
fragment yield in a real nucleus-nucleus collision is still beyond
possibility. For this reason, one is left with the task of trying to gain
understanding of the phenomenon through methods that come from outside the
mainstream of nuclear physics. A number of such models have been devised,
and among them the most successful ones take advantage of statistical
methods and, in particular, of the ideas of universality put forth on firm grounds through the renormalization group approach \cite{wilson}. The statistical multifragmentation models (for a recent review, see \cite{ssm}), for instance, can describe quite satisfactorily the universal properties of multifragmentation. On the other hand, percolation \cite{campi86,bauer} and lattice gas models \cite{latticegas} can also reproduce the inclusive observables, such as the mass or charge distribution, and give support to the interpretation of nuclear fragmentation as some form of critical transition.\\

In the vicinity of criticality, modern theory of phase transitions points
to the irrelevance of many of the fine details that traditional models
usually encompass. One is led, instead, to a search for a correct {\bf
minimal model}, which most economically caricatures the essential physics.
From this point of view, all of the microscopic physics is subsumed into
as few parameters as possible \cite{goldenfeld}.\\

Following these lines, we developed a simple dynamical lattice model that
takes into account some reasonable assumptions about nuclear multifragmentation:\\

\noindent - The entrance and exit channels are decoupled;

\noindent - The fragments come from the decay of a relatively
(thermodynamically) equilibrated, but still highly excited, source; 

\noindent - Decay proceeds through surface rearrangements of the nucleons
that compose the source, and through bulk processes. Fragmentation is
the outcome of surface relaxation.\\

In section 2 we present the model devised with the above ideas in mind and discuss some alternatives for its implementation. In section 3 the results of our simulations are compared to both experiment and previous theoretical results. Conclusions are presented in section 4.
 
\bigskip
\noindent{\bf 2. A Dynamical and Dissipative Lattice Model}
\bigskip

To mimic the essence of the physics of nuclear fragmentation we consider a lattice model and describe it using a terminology that pertains to magnetic systems.\\

The nucleus is represented by a number of connected sites on a hypercubic lattice. A site is occupied by a nucleon {\it iff} it is associated to a spin up. The model dynamics is driven by an Ising Hamiltonian with nearest and next-nearest couplings 
\begin{equation} {\cal H} = - J\sum_{<i,j>} S_{i}S_{j},
\label{H} \end{equation} where $S_{i,j}= \pm 1$.
The inclusion of next-nearest couplings has the purpose of reproducing in a more complete way the energy associated to surface tension; this is highly desirable in our case since our key assumption is that fragmentation comes as a result of surface relaxation.\\

The initial state of the excited nucleus is represented by a 6x6x6 cluster of up spins embedded in a cubic lattice of edge 32 - the initial mass of the nucleus is thus 216; this particular number was chosen so as to easen up straightforward comparisons to experimental and theoretical data available in the literature \cite{campi86, gilkes}. The initial temperature is an input datum, and is translated into excitation energy through one of a number of different strategies - the relation between the two is known in the nuclear physics literature as a caloric curve. We measure the temperature of the excited nucleus in units of the Onsager critical temperature for the 2-dimensional nearest neighbour model. \\

The system is subjected to a Metropolis mass-conserving Kawasaki-like dynamics \cite{kawa}: at each step in the simulation, two non-adjacent sites, one in the inner perimeter and the other in the outer perimeter, are randomly chosen. If it is energetically favorable to flip both spins - thus transfering a nucleon from the inner perimeter to the outer one - it is done; if not, the spin flip is still possible, and is performed with a probability given by a Boltzmann distribution. The energy difference between successive configurations is accumulated.\\

At each spin flip, we determine if the system is still connected; if not, the smallest cluster is considered as a fragment, its mass is included in a distribution for later statistical treatment, and it is erased from memory. The determination of system connectedness is the most time-consuming part of our computer code, and we use the so called "burning" algorithm for that \cite{burn}. The excitation energy is decreased by a quantity proportional to the accumulated energy difference between the actual configuration and the one present when the last fragment was formed. A caloric curve is used to generate the new temperature from the decreased excitation energy.\\

At each fixed number of steps - in our case, 100 - the excitation energy
is decreased by a fixed factor $a$ - the model's only parameter - representing bulk radiative decay. The combination of this parameter and the above mentioned number of steps can be thought of as a relation between the time scales of the surface and bulk decay rates.\\

The simulation proceeds until the residual nucleus has too small a mass or
a temperature, whichever comes first. We chose a minimum mass of 2 and a minimum temperature of 0.99 of Onsager's temperature for these values. For each value of the initial temperature $2000$ events are simulated, with different seeds for the random number generator. A typical run takes approximately 8 hours on a 433 DEC Alpha workstation.\\

For the caloric curve, we used a linear or a Fermi-gas type - quadratic -
relation. The results of the simulation are the fragment distribution and
its various moments.\\

A very similar model has already been used in different contexts to study
wetting \cite{landau}, drop formation on the leaky faucet problem
\cite{thadeu} and, in nuclear physics problems, to reproduce the
distribution of mass of the remnant nucleus in evaporation processes
\cite{evaporacao} and to put in evidence the relation between moments of
different orders of the fragment distribution \cite{modelo}.

\bigskip
\noindent{\bf 3. Results of the Simulations}
\bigskip

We have run our simulations with a number of different values for the ratio between the time scales for surface and bulk relaxations and two different caloric curves, as already described. Nuclear multifragmentation is believed to be in the percolation universality class \cite{universal}, and we could obtain the correct critical exponents with a Fermi gas type caloric curve and a model parameter $a=0.998$. Different values for this parameter produce also different exponents, and this is a somewhat surprising characteristic of the model that will be better investigated in the future.\\

We adress at first the question of how to measure the distance to the critical point. There is no model-independent way to choose the appropriate physical parameter to use. Although temperature is a natural choice, its inaccessibility in nuclear collisions experiments forces one to consider a different quantity. There is general agreement on the fact that the multiplicity of the event - the number of intermediate mass fragments - is linearly related to the temperature \cite{hauger}, at least next to the critical point, and this latter quantity is the one frequently chosen by the experimentalists.

From Fig. 1 we can see that, to a good approximation, our model produces a linear relation between multiplicity and temperature in a region of temperature values that encompass the critical point - which will be shortly shown to be $T_{c}=8.50$ - if a Fermi gas caloric curve is used. This gives additional support to the experimentalist's choice, and allows a direct comparison between our results and experiment.\\

Through an analogy between fragment formation and percolation, one is led
to study how the size - mass - of the largest fragment evolves in the
vicinity of the critical point. For an infinite system, this quantity is
the order parameter of the transition - the infinite percolating cluster. 
Fig. 2 shows the dependence on temperature of the relative size of the
largest fragment, and in the inset we can see its critical fluctuations. This quantity clearly has the correct features for a candidate to an order parameter for the fragmentation critical behaviour. In practice, it is through the analysis of its fluctuations that the critical region is identified.\\

The critical behavior of nuclear multifragmentation as a remnant of the liquid-gas phase transition in infinite nuclear matter can be put in evidence through the linear relation between the logarithms of the third and second moments of the fragment distribution \cite{campi86}. The slope of this relation $\lambda_{3/2}$ is connected to the exponent $\tau$ of the scaling relation \cite{fisher, stauffer} $$n(s,\epsilon) \sim s^{-\tau}f(\epsilon s^{\sigma}),$$ where $\epsilon$ is the reduced distance to the critical point, $s$ is the mass of the fragment, $n$ is its yield and $\tau$ and $\sigma$ are two critical exponents , by $$\tau = 3 - \frac {1}{\lambda_{3/2} - 1}.$$ Our model reproduces the experimental and 3D percolation results \cite{modelo}, with $\tau=2.18$.\\

To determine a second critical exponent we use a method devised in Ref.\cite{gilkes} for $\gamma$, the exponent of the second moment of the fragment distribution $$M_2(T) = \sum {n(s,T)s^2}.$$ In analogy with percolation, the largest fragment is
included when computing this moment in the gas phase, but not in the liquid phase, where it plays the role of the infinite percolating cluster. These moments are then fitted to a critical behavior $$M_2 \sim \left\vert T-T_c \right\vert ^\gamma,$$ at both sides of the transition, and $T_c$ is determined through the identity of the exponent $\gamma$ calculated in the liquid and in the gas phases.

In Fig. 3 we show our results for this fitting. For a critical temperature of 8.50, we could obtain $\gamma=1.80$ as in 3D percolation.\\

\bigskip
\noindent{\bf 4. Conclusions}
\bigskip

We could obtain with a dynamical and dissipative lattice model, for a suitable choice of the caloric curve and the ratio between the time scales for surface and bulk relaxation - namely, a Fermi gas relation between
excitation energy and temperature and a constant decremental factor for
excitation energy of $a=0.998$ - the same $\tau$ and $\gamma$ exponents of
percolation. Since there is some experimental evidence pointing to this
as the universality class of the multifragmentation process, we believe
that this shows that the physical ingredients of our model can be of importance for a better understanding of that complex phenomenon. In particular, the introduction of two different time scales for the competing processes of surface and bulk relaxation turned the model able to be tuned up to a number of sets of critical exponents.

\bigskip 
\noindent We acknowledge the Brazilian agencies CNPq, CAPES,
FAPERJ and FINEP for financial support. 

\bigskip

\newpage
\centerline{\bf Figure Captions}
\bigskip

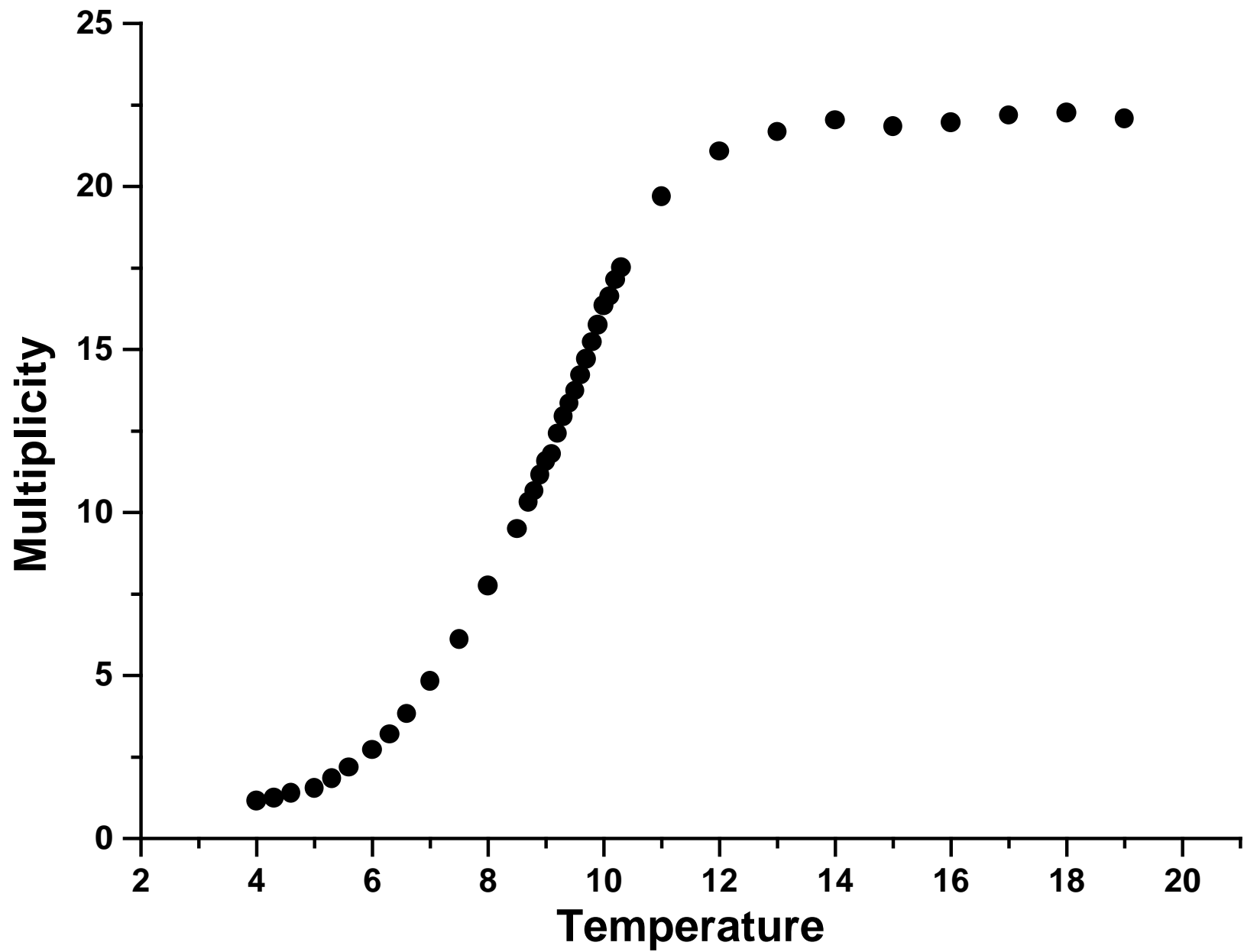
\noindent {\bf Fig.1} - Mean multiplicity - number of intermediate mass fragments - as a function of initial temperature, measured in units of the critical temperature of the Ising 2D model. Error bars, computed through the variance of the distribution, are of the order of the size of the symbol used and are not shown. All the results come from simulations in 3D, for a Fermi gas caloric curve and a value of $a = 0.998$.
\bigskip

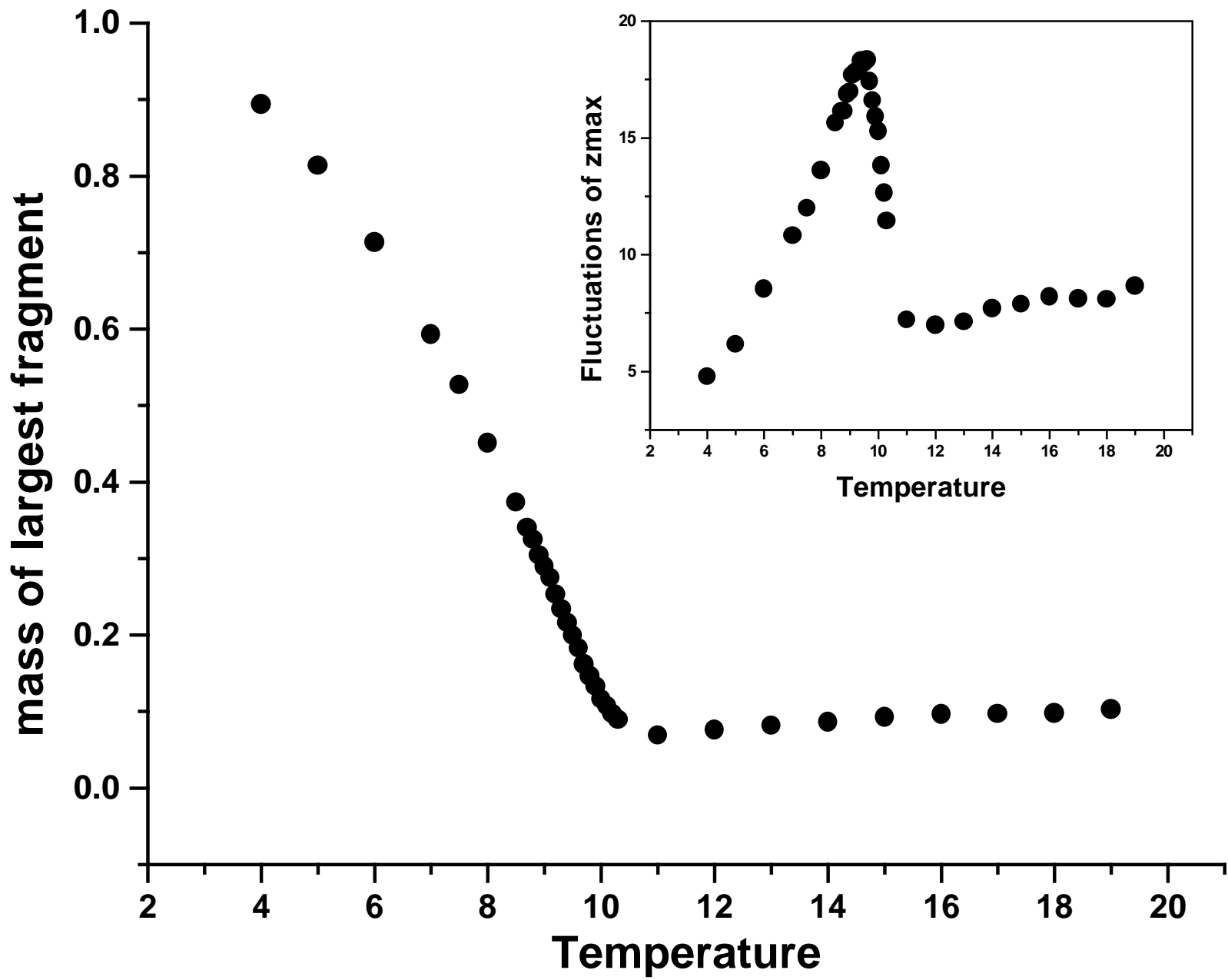
\noindent {\bf Fig.2} - Mean mass of the largest fragment as a function of initial temperature. The inset shows the fluctuations of the mass of the largest fragment, in arbitrary units, as a function of the initial temperature. 
\bigskip

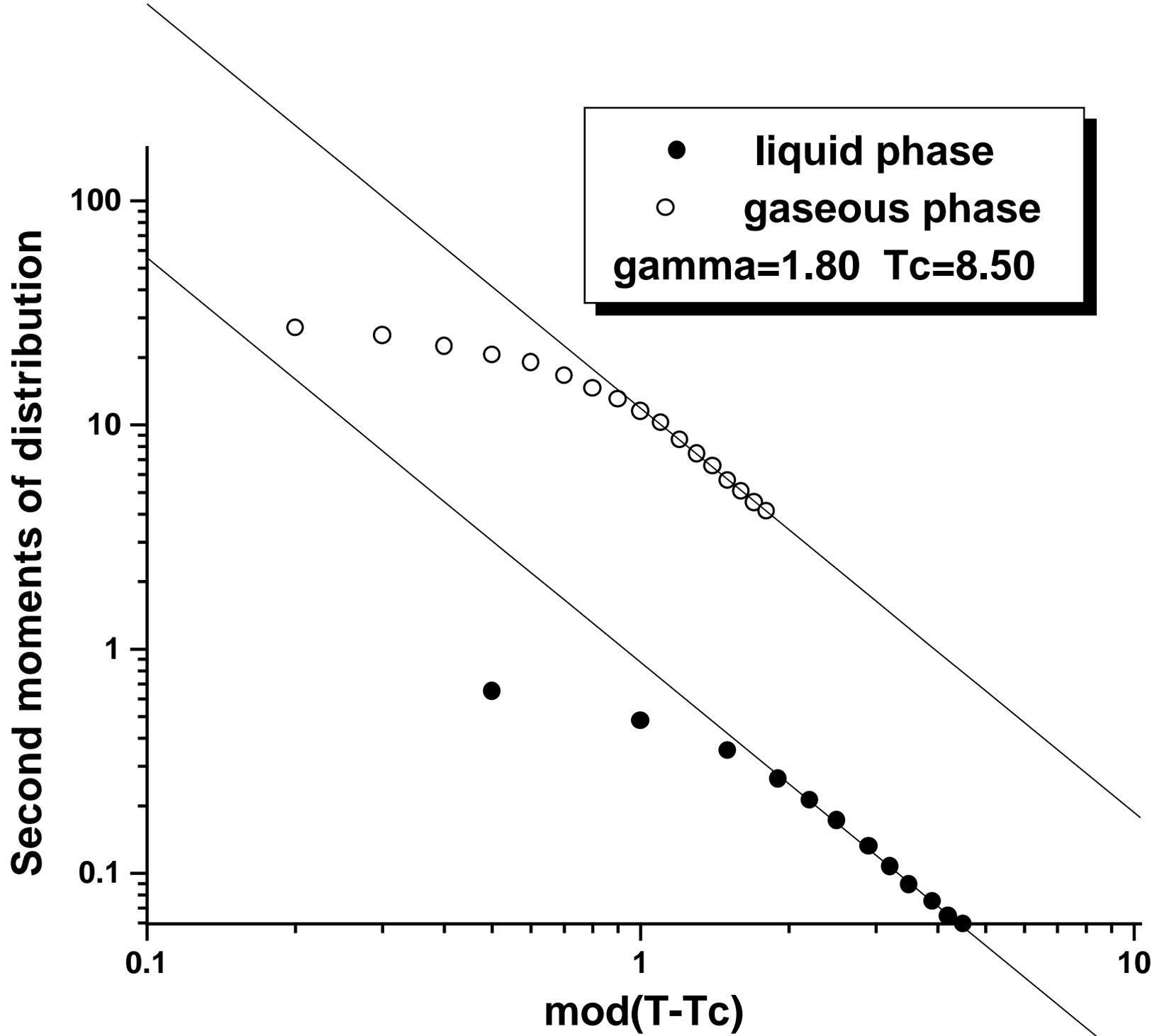
\noindent {\bf Fig.3} - Second moments of the fragment distribution at both sides of the fragmentation transition, as a function of the distance to the critical temperature, in a log-log plot - see text. The power law fits shown determine the critical temperature and the $\gamma$ exponent. 
\bigskip

\end{document}